\documentstyle[11pt,epsf]{article}
\begin{document}
\epsfclipon

\title{\bf Experiments on the generation of long wavelength edge
radiation along directions nearly coincident with the\\  axis of a
straight section of the "Pakhra" synchrotron } \author{V.I.Alexeev,
E.G.Bessonov} \date{} \maketitle

\begin{center}
{\it Lebedev Physical Institute of the Russian Academy of Sciences,
Moscow, Russia}
\end{center}

\abstract{ Generation of long wavelength edge radiation along
directions nearly coincident with the axes of straight sections of
storage rings and synchrotrons is discussed. The removal of destructive
interference from the superimposed edge radiation patterns of a straight
section on the synchrotron "Pakhra" has been observed experimentally.}

                   \section{Introduction}

The concept of long wavelength radiation was introduced in
classical electrodynamics for the case of a particle emitting radiation
in external fields. The spectral-angular distribution of the long
wavelength electromagnetic energy emitted by the particle doesn't
tend to zero when the frequency of the emitted radiation tends to zero
\cite{landau, jackson}. The electromagnetic wave components of
long wavelength radiation have mainly one sign. The Fourier transform
of their electric field strength $\vec E_{\omega}|_{\omega = 0 } \ne
0$ \cite{bes1} - \cite{bes4}.

The object of this paper is to discuss the problem of generation of
radiation with properties near to the long wavelength radiation
by a particle or particle beams in external electromagnetic
fields of synchrotrons and storage rings.  Some experimental
results carried out in the beginning of 1980 will be presented.

\section{Selected questions of the classical electrodynamics}

The electric and magnetic field strengths of a non-uniformly moving
charged particle are determined by

                  \begin{equation}   
          \vec E(t) = \vec E^c(t) + \vec E^r(t),
          \hskip 5mm \vec H = [\vec n \vec E],
                     \end{equation}
where
          $$\vec E^c(t) = {e(1-\beta ^2)(\vec n-\vec \beta)
          \over R^2(1-\vec n\vec \beta)^3}|_{t^{'}}, \hskip 15mm
          \vec E^r(t) = {e[\vec n[(\vec n-\vec \beta )\dot {\vec \beta
          }]]\over cR(1-\vec n\vec \beta)^3}|_{t^{'}},$$
e, c$\vec \beta$ and c$\dot {\vec \beta }$ are the charge, velocity, and
acceleration of the particle; $\vec n$, the unit vector directed from a
particle to the observation point, and $R$, the distance from the
particle to the observation point \cite{landau},\cite{jackson}. The
values $\vec \beta$, $\dot {\vec \beta }$, $\vec n$ and $R$ are to be
evaluated at an earlier (retarded) moment $t^{'} = t - {R(t^{'})/c}$.
The terms $\vec E^c(t)$, $\vec H^c(t)$ in Eq1 describe the sharply
decreasing ($\sim 1/R^2$) Coulomb field of the particle, while
$\vec E^r(t)$ and $\vec H^r(t)$ are for the free electromagnetic
field radiated by the particle ($\sim 1/R$).

Usually the observation point is assumed to be far away from the region
of particle acceleration, the unit vector $\vec n$ is sensibly
constant in time \cite{landau, jackson}. The region of particle
acceleration and emission is assumed to have finite dimensions, and the
radiation is observed from distances much larger than dimensions of the
emission region (in the far-zone or wave zone). The particle emits
radiation within a finite time interval determined by initial and
final moments $t_i$, $t_f$ of particle acceleration. The energy of the
emitted radiation is finite.

The properties of the emitted radiation are defined by the Fourier
transform $\vec E_{\omega}^r$ of the electric field strength vector
$\vec E^r(t)$ of the form

           \begin{equation} 
           \vec {E_\omega ^r} = {1\over 2\pi}\int _{- \infty}^{+
           \infty}\vec E^r(t) e^{i\omega t}dt.
           \end{equation}

For instance, the spectral-angular distribution of the energy radiated
by a particle into the solid angle $do = dS/R_o^2$ or onto the area
$dS$ at the observation point is determined by

           \begin{equation}  
           {\partial ^2\varepsilon \over \partial \omega
           \partial o} = R_o^2{\partial ^2\varepsilon \over \partial
          \omega \partial S} = cR_o^2|\vec E_{\omega}^r|^2.
           \end{equation}

According to Eq2 the properties of the emitted radiation are
determined by the total trajectory of the particle where the particle
acceleration differs from zero. This assertion is in accordance with the
Fourier transform of Eq2 for the time interval ($-\infty,
+\infty$) or for the real case ($t_i, t_f$) \cite{bes4}.

Notice that the Coulomb fields have longitudinal component and $|\vec
H^c| \ne |\vec E^c|$ ($|\vec H^r| = |\vec E^r|$). These are the reasons
why the Fourier transform of the Coulomb and total electric and
magnetic fields in near zone could not be introduced and interpreted
in a simple way. Only in the ultrarelativistic case, where $|\vec H^c|
\to |\vec E^c|$, the Fourier transform of the Coulomb field is
interpreted as the superposition of the equivalent elementary waves
corresponding to virtual quanta in the Weizsacker-Williams method. But
this is the approximate method used only for the case of homogeneously
moving particles in empty space. It describes well the bremsstrahlung
radiation of electrons in the fields of relativistic ions
\cite{jackson}. In other cases the accuracy of this method is not
high. For example, when the relativistic particle crosses a plane mirror
it emits the transition radiation. We can say that in this case the
properties of the transition radiation are defined by the properties of
equivalent photons "reflected" from the mirror. However, the
properties of the transition radiation are calculated precisely and they
differ from those calculated by the method of
equivalent photons. For instance, in the case of a perfect mirror the
transition radiation fields are included in the spherical layer of the
thickness $d\to 0$ (the layer contains the emitted radiation of
wavelengths $\lambda \gg d$ ). They are emitted from one point or from a
very small region \cite{jackson} (equivalent photons would be reflected
by the total surface of the mirror), and have a similar
spectrum \cite{bes-shib}.

In some cases, the total fields (distorted Coulomb and radiation fields)
of a particle must be taken into account, e.g. a distortion of
scattered particles fields can change the ionizing properties of
the particles \cite{bes1}. At the same time, the inclusion of the
Coulomb fields in the Fourier transform in some papers \cite{sokol},
\cite{hirai}, \cite{mathis} is not motivated and does not change
essentially properties of the emitted radiation \cite{castellano}. Of
course, the account of retarded Coulomb near-fields of high current
beams in the storage rings \cite{iogansen}, \cite{murphy},
\cite{melechin}, \cite{carlsten} and in undulators of free-electron
lasers \cite{bes4}, \cite{saldin} is necessary to obtain the correct
particle trajectories and to calculate the total losses of the particle
energy through coherent radiation \cite{bes4}.

The Fourier transform of the electric field strength at zero frequency
can be presented in the form $\vec E_{\omega}^r|_{\omega = 0} = (1/2\pi
)\vec I^r$, where $\vec I^r = \int _{-\infty}^{+\infty}\vec E^r(t)dt$ is the
strange parameter of the wave. Waves with this parameter $\vec I^r
\ne 0$ were named strange waves. Strange waves have components of the
electric field strength mainly of one sign. In particular, single-sign
waves are the strange waves \cite{bes1}.

Strange waves transfer to a charged particle a momentum in the
direction transverse to the wave propagation. This
property of strange waves contradicts to the common opinion that light
pressure is directed only in the direction of light propagation. The
question appears:"Do the strange waves exist? Is it possible to
generate them?" The answer is not trivial.

When they say the word "radiation" it usually means free
electromagnetic waves described by a homogeneous wave equation.  The
solution of this equation can be the arbitrary function of argument
$(\vec n\vec R - ct)$ \cite{landau, jackson}. The word "arbitrary"
means that single-sign waves (strange waves) satisfy the
homogeneous wave equation in particular case. However, the existence of
such solutions is a necessary but not a sufficient condition for the
existence of strange waves. What is the sufficient condition? To answer
the question we must solve the nonhomogeneous wave equation.

In the case of one particle the solution of the nonhomogeneous wave
equation is described by the Lienard-Wiechert fields of Eq1. Integration
of the electric field strength $\vec E^r(t)$ determined by Eq1
leads to the strange parameter of the free field emitted by the
particle

                   \begin{equation} 
               \vec I^r = {e\over cR_o}[\vec n [\vec n({\vec \beta _2\over
             1-\vec n\vec \beta _2} - {\vec \beta _1\over 1 - \vec n
             \vec \beta _1})]], \end{equation}
where subscripts $1,2$ relate to initial and final electron velocities
\cite{bes1}.

According to the Eq4 strange waves can be emitted only in the case when
the initial and/or final velocities of the particle are not equal to
zero and hence the particle trajectories are not limited.  Unacceptable
conditions for particle trajectories exist for laboratory sources of
strange waves as according to these conditions the dimensions of such
sources must be infinitely large. It is possible to emit strange waves
in cosmos but impossible to emit them in installations of finite
dimensions similar to lasers or synchrotron radiation sources
\cite{bes1}.

\section{On the laboratory sources of the long wavelength radiation
based on the storage rings and accelerators}

At present the high brilliance and high intensity broadband sources of
both incoherent and coherent infrared radiation are absent. The long
wavelength part of the synchrotron and edge radiation \cite{bes1} -
\cite{bes4} emitted in the infrared region by particle beams in storage
rings is increasingly used in different fields of science as an
alternative for laboratory sources such as a mercury lamp or glowbar
\cite{schweizer}, \cite{mathis}.  Broadband free-electron lasers of the
coherent infrared radiation are developing \cite{murphy2}, \cite{hung},
\cite{sendai}.

The efficiency of emission of the synchrotron radiation in the long
wavelength region in storage rings and synchrotrons is increased when
the value of the magnetic field strength of their bending magnets is
decreased.  Usually the value of the magnetic field strength of bending
magnets of storage rings is high.  That is why we can use special
bending magnets installed in the straight sections of storage rings.
The value of their magnetic field strengths have an optimum defined by
the equation $e\int Hdl \simeq 2mc^2$ corresponding to the bending
angle of the particle in this magnets $\simeq 2/\gamma$, where the
value $2mc^2/e \simeq 3 \cdot 10^3$ Gs$\cdot$cm and $\gamma $ is the
relativistic factor of the particle.  First experiments confirmed this
statement \cite{artem}.  The radiation from special bending magnets in
such cases interferes with the edge radiation. As it turned out,
at optimal conditions of generation the magnetic field introduced
in the straight section must be directed opposite to the
guiding magnetic field of the storage ring. In the general case, the
system of bending magnets installed in the straight sections of the
storage rings can produce both linear and circular polarized broadband
radiation in the long wavelength region \cite{bes2}.

In the case of pure edge radiation (special bending magnets in the
straight section of the storage ring are absent) the components of the
electric field strength of each of the two wavepackets emitted in the
edge fields of the storage ring at the angle $\sim 1/\gamma$ to the
direction of the axis of the storage ring have mainly one sign.  But
signs of wavepacket components are opposite. It means that every of the
wavepackets represents the strange wave radiation. Its spectrum does not
tend to zero when $\omega \to 0$. However the sum of the wavepackets is
not the strange radiation because of the destructive interference. The
lengths of the wavepackets in this case are much less than the distance
between them.  The picture of the process in space is similar to the
case of an instant start and instant finish of a particle at
the edges of the storage ring. The edge radiation can be named
"conventionally strange" or "quasiundulator" radiation \cite{bes2} -
\cite{bes4}\footnote{When the difference between the distances from the
edges to the observation points are of the order of the effective
length of the straight section of the storage ring then the strange
parameter $I^r = |\vec I^r| \ne 0$ if we neglect boundary conditions.
However the windows of synchrotrons and mirrors of finite dimensions
do not permit to transfer this radiation to the observation point
\cite{bes1}.}.

The spectral-angular distribution of the edge radiation emitted in the
long wavelength region $\lambda > d/2\gamma ^2$ and under the condition
when the observation point is localized at infinity (condition when we
can neglect the difference between distances from edges of the
synchrotron to the observation point) can be presented in the form

           \begin{equation} 
           {\partial ^2\varepsilon \over
           \partial \omega \partial o} = {4e^2\over \pi ^2
           c}{\vartheta^2\over (1 + \vartheta ^2)^2}\sin ^2{\omega
           \over \omega _l}, \end{equation}
where $l_{eff}$ is the effective length of the straight section of the
storage ring, $l - l_{eff}|_{H \gg H_d} \simeq 2d \ln (H/H_d)$; $l$, the
geometrical length of the straight section, $d$, the gap of the bending
magnet of the storage ring, $H$, the value of the magnetic field of the
bending magnet of the storage ring, $H_d = mc^2/ed$; $m$, the electron
mass, $\vartheta = \gamma \theta$, $\omega _l = 4\gamma ^2c/l _{eff}(1
+ \vartheta ^2)$, and $\theta$, the angle between the axis of the
straight section and the direction from the synchrotron edge to
the observation point localized at infinity. The effective length
$l_{eff}$ and the frequency $\omega _l$ depend slightly on
energy \cite{bes2}.

Spectrum of the edge radiation is shifted to the long wavelength
region. In this case the spectral-angular distribution of the radiation
emitted at $\theta \sim 1/\gamma$ is higher than that of the
synchrotron radiation (for the same synchrotron) \cite{bes-shib}.

According to Eq5 the main part of the edge radiation energy is emitted
throughout the angles $\vartheta \sim 1$. At frequencies $\omega >
\omega _{l}$ the spectrum oscillates with the period $\omega _l$ and at
frequencies $\omega < \omega _{l}$ it tends to zero when the frequency
tends to zero, which indicates that the wavepackets have opposite
signs. This is the so-called destructive interference of the
wavepackets of the opposite signs first discussed in the case of
transition radiation\footnote{The transition radiation emitted in the
case of unlimited perfect plane mirror is the strange wave radiation.
}\cite{wartsky}.

\section{Experimental arrangement}

The experiments on generation of the long wavelength edge radiation
were done on the weak-focusing race-track electron synchrotron
"Pakhra".  The maximum energy of the synchrotron $\varepsilon _{max} =
1200 MeV$, the radius of the orbit of electrons in the bending magnets
of the synchrotron $r = 4 m$, the geometric length of a straight
section of the synchrotron $l = 1.9 m$. The frequency of the guiding
magnetic field of the synchrotron $\Omega/2\pi = 50$ Hz.

The experimental setup of the "Pakhra" synchrotron used in the recent
experiment \cite{alex}, \cite{al} is similar to that presented in
\cite{alferov}. The scheme of the experiment is shown in
Fig.1, where $l$, $l _{eff}$, the geometrical and effective lengths
of the synchrotron straight section, $R_1$ and $R_2$ the
distances from edges of the synchrotron straight section to the
observation point, $\theta _1$, $\theta _2$ the angles between the
directions of the axis of the synchrotron straight section and vectors
directed from edges of storage rings to the observation point.

The energy of electrons in the synchrotron was changed by the law
$\varepsilon _e = \varepsilon _{=} + \varepsilon _{\sim}\cos(\Omega
t)$, where the constant and alternating components of the energy
$\varepsilon _{=} \simeq \varepsilon _{\sim}$. The electron energy
changed from minimal $\varepsilon _{min} = \varepsilon _{=} -
\varepsilon _{\sim}$ to a maximal $\varepsilon _{max} =
\varepsilon _{=} + \varepsilon _{\sim}$. The maximal energy of the
electrons in this experiment was chosen $850$ MeV ($\gamma \simeq
1700$).

The scheme of selection of the wavepacket emitted from one edge of
the synchrotron by a lens having the focusing distance $f$ and located
at the distance $L$ from the right edge is shown in Fig.2. In two of
produced experiments, $l_{eff} \simeq 1.25 $ m and $L = 1.875$
m. The lens was installed at the window of the synchrotron,
photographic plate in the first experiment and photomultiplier in the
second were installed at distance $\Delta R =1.5$ m from
the lens. Filters at the wavelengths $\lambda = 8400 \AA$ and $\lambda =
3850 \AA$ and lenses with focusing distances $f = 1$ m  and $f = 1.85$
m were used in the first and the second experiments.

The observation point in our experiments was localized at the distances
$R_1 = 4.625$ m and $R_2 = 3.375$ m. The difference $R_1 - R_2 =
l_{eff} = 1.25$ m was about 3 times less than $R_1$ and $R_2$ and could
therefor be neglected in the experiments without the lens. In fact it
could be taken into account by introducing in Eq4 unit vectors
$\vec n_1$ $\vec n_2$ and distances $R_1$, $R_2$ corresponding to the
edges of the bending magnets of the synchrotron \cite{bes1}. The effect
of using of the lens is the higher the higher the difference between
$R_1$, $R_2$ and $l_{eff}$.

\section{Discussion}

According to Eq4, the condition $\beta _1 = \beta _2 = 0$ for the
electron trajectories in the synchrotron, and the definition presented
in the Introduction, the radiation emitted in the synchrotron is not the
long wavelength one. The radiation emitted from the central part of the
bending magnet of the synchrotron in the direction near to the
direction of the electron velocity is a usual synchrotron radiation.
In the direction of the corresponding observation point this radiation
is emitted mainly from the part of the trajectory of the length $\sim
l_{c} = 2R/\gamma$.  The edge radiation is emitted from the adjoining
edge regions of fringing fields of bending magnets.  In this case the
components of the electric field strength of the electromagnetic waves
emitted in the directions defined by the angles $\theta \sim 1/\gamma$
are the two short ($l _s \ll l_{eff}/\gamma ^2$) single-sign wavepackets
separated by the distance $\sim l_{eff}/\gamma ^2$. Spectrum of the
edge radiation determined by Eq5 is shifted to the long wavelength
region, as compared to synchrotron radiation spectrum.

If we could select the radiation emitted in one bending magnet from the
radiation emitted in another one then we could shift the emitted
radiation in a longer wavelength region. And this is not the
question of the strange waves generation but the question of the
spectrum shift\footnote{To reflect strange waves we need in perfect
plane mirror of infinitely large dimensions. The mirror of finite
dimensions can reflect radiation up to the wavelength $\lambda \sim a$
where a is the dimensions of the mirror. The window permit to transfer
the radiation shorter then the diameter of the window \cite{bes1}.}.

In the first experiment the lens was not installed. No interference
fringes were observed at a photographic plate similar to those
observed in \cite{wartsky, nikitin} since the picture
was averaged over time (energy). When the lens was installed then the
radiation emitted from the left edge was focused onto a small area at
the left side of the picture. The picture of the radiation focused by
the lens is presented in Fig.3. Similar picture was observed at
the distance $\Delta R = 2.3$ m from the lens where the radiation
emitted from the right edge of the synchrotron was focused onto a small
area. So the radiation emitted in the region of one edge
of the synchrotron can be focused and hence selected by diaphragms from
the radiation emitted in the region of another edge.  We did not
observe any contribution of radiation emitted from central part of
the straight section of the storage ring and defined by the Coulomb
term of Eq1.

In the second experiment, the time dependence of the intensity of the
emitted radiation was measured by a photomultiplier and recorded by the
oscillograph. This dependence is shown in Figs. 4 and 5.

The signal of Fig.4 corresponds to the case where the radiation is not
focused by lens. In accordance with the theory (see Eq5) the
interference of the wavepackets emitted from edge regions of the
synchrotron leads to the energy dependence of the frequency $\omega
_l$ and intensity of the emitted radiation on energy and, hence, time.
This observation is confirmed by Fig.4. The signal measured by the
photomultiplier and recorded by the oscillograph oscillates in time.

In Fig.5, the signal is caused by the radiation which is focused by
a lens. We can see again that the lens destroys the interference and
selects the radiation emitted from the region of one edge of the
synchrotron.

Selection of the radiation emitted in one edge of the bending magnet of
the synchrotron permits one to hope that the radiation emitted at the
angles $\theta \sim 1/\gamma$ to the axis of the synchrotron will be
shifted to the longer wavelength region.

It is impossible to focus the radiation at all frequencies by a glass
or silicon lens. Metal concave mirror can be used in optical, IR and
submillimeter regions. Dimensions of windows of synchrotrons and
storage rings and dimensions of mirrors restrict the spectrum from the
long wavelength region.

\section{Conclusion}

For the first time, to our knowledge, the removal of the destructive
interference in the edge radiation has been demonstrated here
experimentally. We can hope that the spectrum of the radiation emitted
in the angular region $\theta \sim 1/\gamma$ to the axis of the
straight section of the synchrotron was shifted to the more longer
wavelength region. This conclusion must be verified.

\vskip 50mm
Presented to the Proc. Int. Symposium "Radiation of
relativistic electrons in periodical structures", Sept.13-16, 1999,
Lake Baikal, Russia (will be published in Nucl. Instr. Meth.).

\newpage
\vskip 1cm
\begin{figure}[hbt]
\centerline{\leavevmode\epsfxsize=140mm\epsfbox[40 377 529 549]{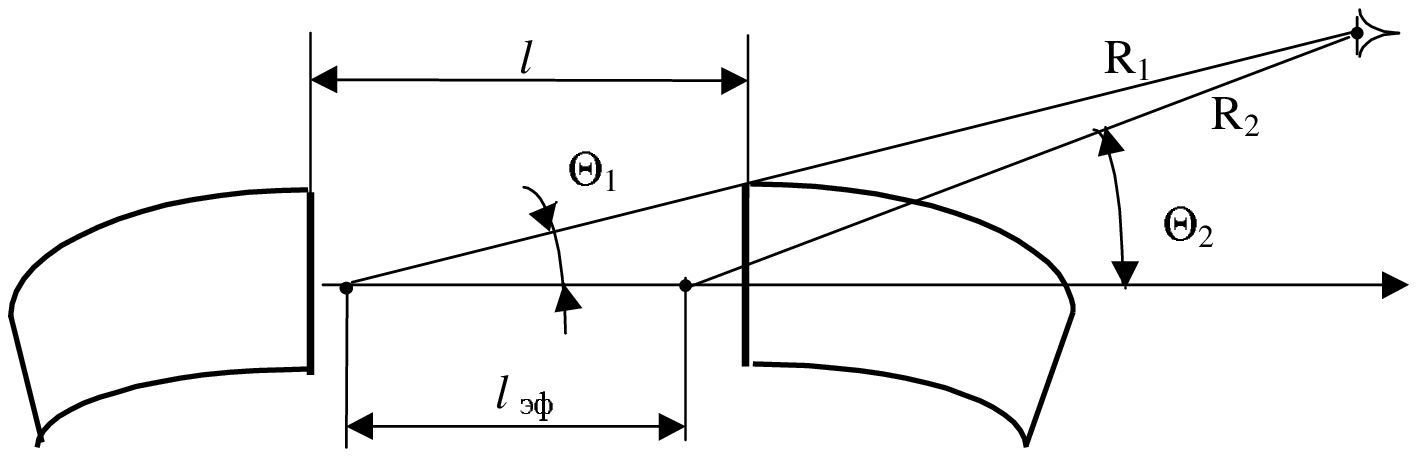}}
\caption{The scheme of the experimental setup.} \end{figure}

\vskip 1cm
\begin{figure}[hbt]
\centerline{\leavevmode\epsfxsize=140mm\epsfbox[70 440 540 600]{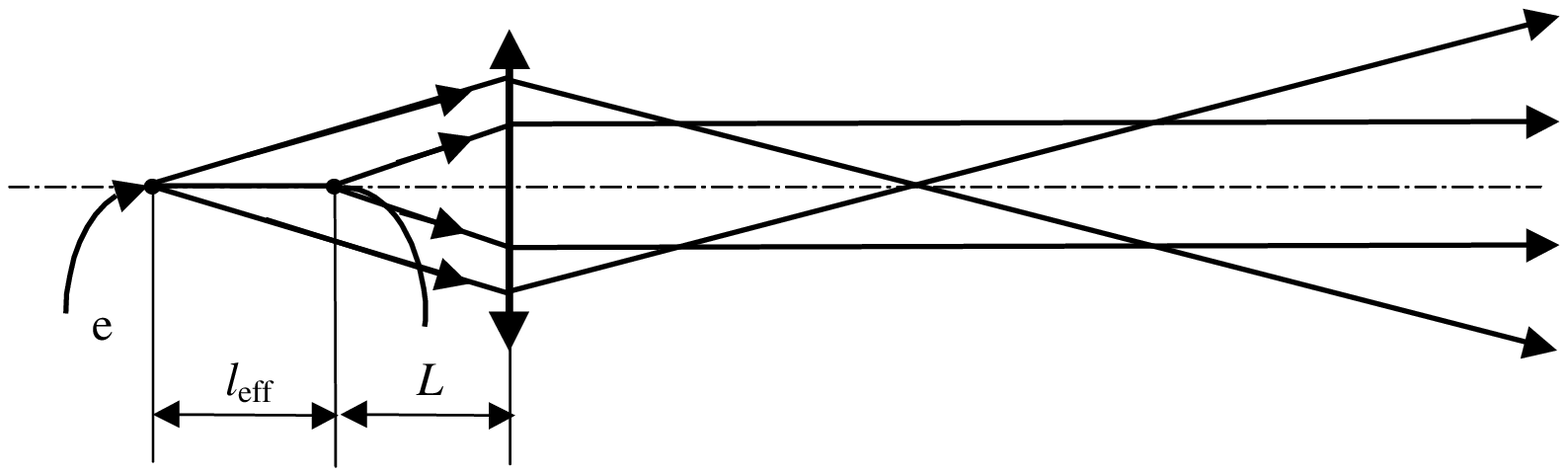}}
\caption{The scheme of selection of the wavepackets emitted from the
right edge of the synchrotron.} \end{figure}

\vskip 1cm
\begin{figure}[hbt]
\centerline{\leavevmode\epsfxsize=140mm\epsfbox[125 377 460 470]{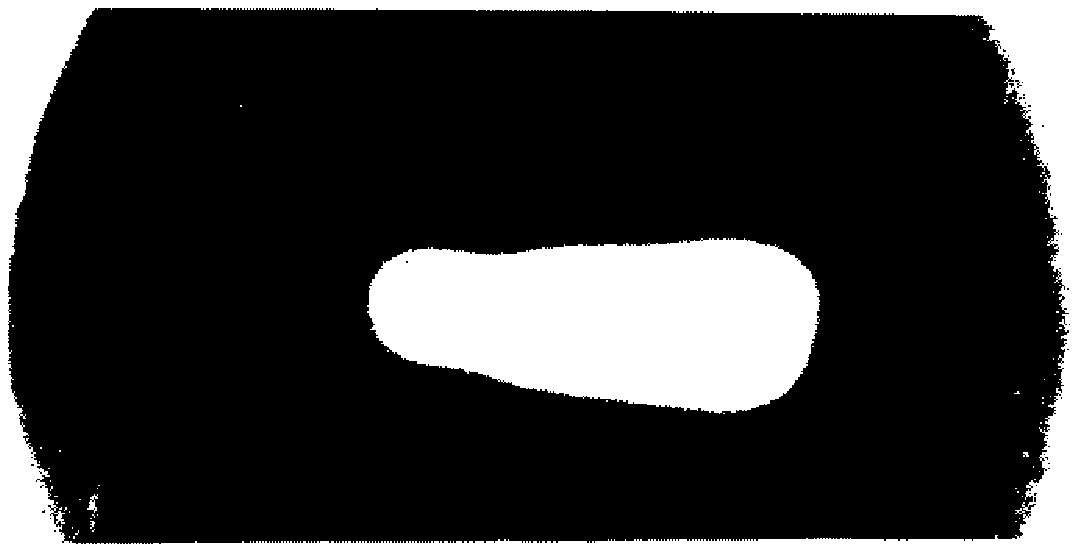}}
\caption{The picture of the radiation emitted in the direction of
straight section of the synchrotron \protect \\ and focused by a lens.}
\end{figure}

\vskip 1cm
\begin{figure}[hbt]
\centerline{\leavevmode\epsfxsize=100mm\epsfbox[83 210 510
500]{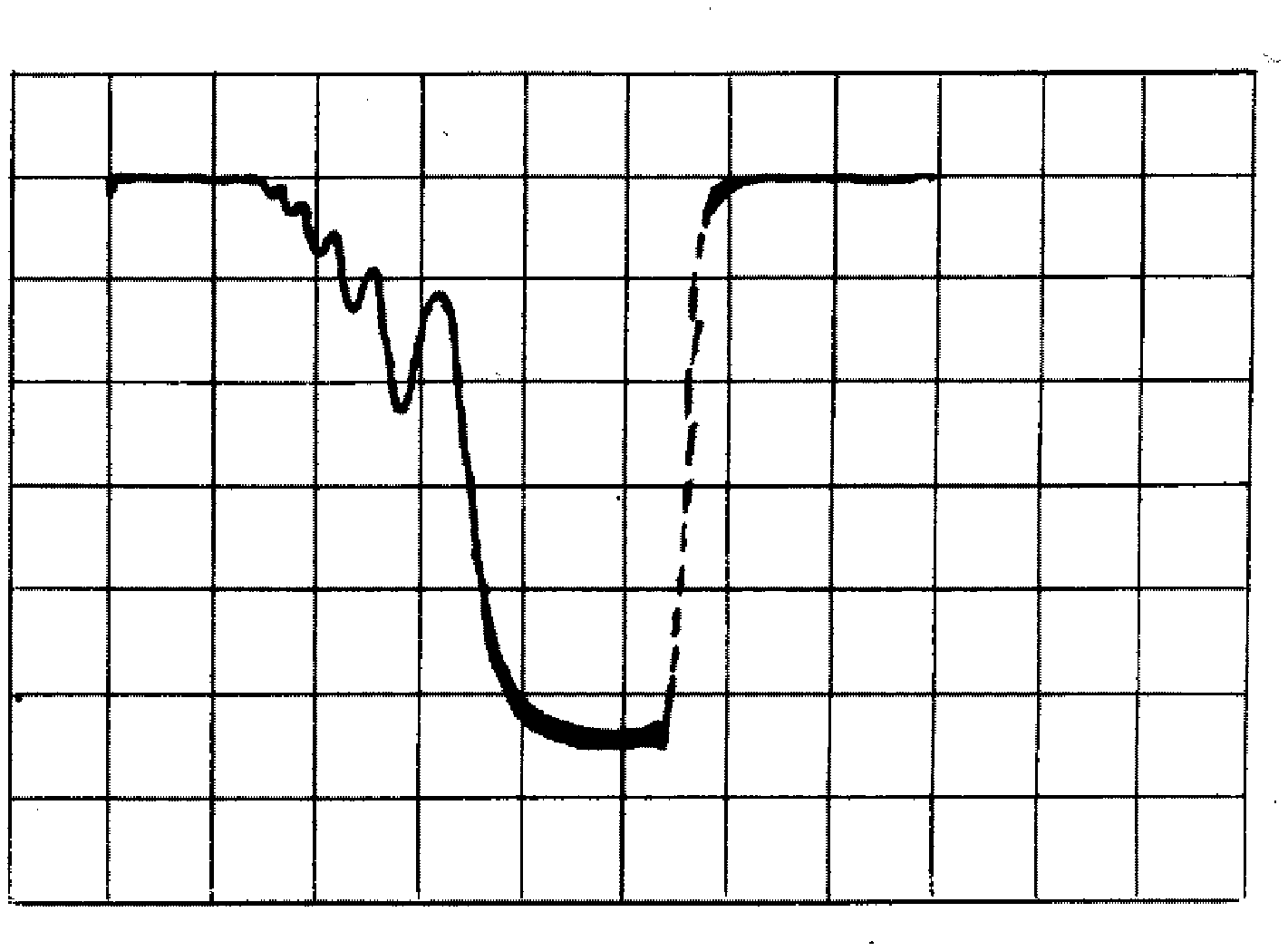}}
\caption{Time dependence of the intensity of light emitted in the
direction of the axis of \protect \\ straight section of the
synchrotron. Focusing lens is absent.} \end{figure}

\vskip 1cm
\begin{figure}[hbt]
\centerline{\leavevmode\epsfxsize=100mm\epsfbox[90 220 500
510]{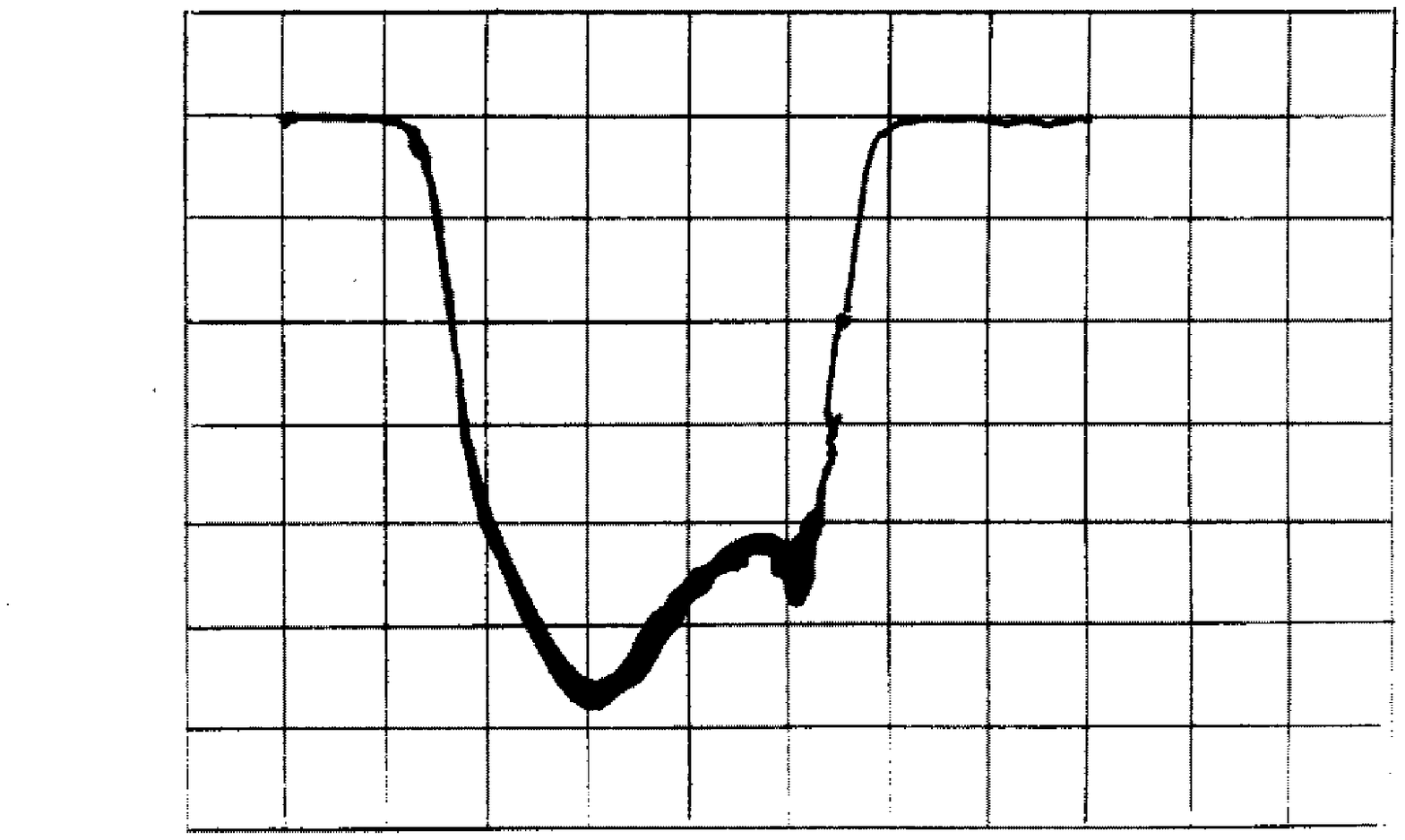}}
\caption{Time dependence of the intensity of light emitted in the
direction of the axis of \protect \\ straight section of the
synchrotron. Focusing lens is installed.} \end{figure}


\begin{thebibliography}
\noindent
\bibitem {landau}  
Landau, L.~D., and E.~M.~Lifshitz, {\it The Classical Theory of
Fields,} 3rd Reversed English edition, Pergamon, Oxford and
Addison-Wesley, Reading, Mass. (1971).
\noindent
\bibitem {jackson} 
J.~D.~Jackson, {\it Classical Electrodynamics,} John Wiley $\&$.  Sons,
1975.
\noindent
\bibitem{bes1}       
E.G.Bessonov, Sov. Phys. JETP 53 (3), 1981, p.433.
\noindent
\bibitem{bes2}         
E.G.Bessonov, Sov. Phys. Tech. Phys., 1983, v. 28 (7), p.837.
\noindent
\bibitem{bes3}         
a) E.G.Bessonov, Nucl. Instr. Meth., 1991, v. A308, p. 135; p. 142,
b) E.G.Bessonov, Proc. 15th Int. Accel. Conf. on High Energy
Accelerators, v.1, Hamburg, Germany, 1992, p. 503; p.501,
c) E.G.Bessonov, M.L.Vnukova, Quantum Electronics, v.25 (12), p.1214,
1995 (English version of Kvantovaya Electronica (in Russian), v.22
(12), 1251 (1995).
\noindent
\bibitem{bes4}       
E.G.Bessonov, Proc.  Lebedev Phys. Inst.  of the RAS, ser.214
(Undulator Radiation, Free-Electron Lasers), Ed. N.G.Basov,
Nauka, 1993, p.3.
\noindent
\bibitem{bes-shib}     
E.G.Bessonov, Yu.Shibata, Preprint FIAN No 35, 1996; physics/9708023.
\noindent
\bibitem{sokol}     
A.A.Sokolov, D.V.Gal'tsov, M.M.Kolesnikova, Izvestia VUZov, No 4,
1971, p.14.
\noindent
\bibitem{hirai}     
Y.Hirai, A.Luccio, L.Yu, J. Appl. Phys. v.55, p.25 (1984).
\noindent
\bibitem{mathis}     
Y.-L.Mathis, P.Roy, B.Tremblay, et al., Phys. Rev. Lett., 1998,
v.80, No 6, p. 1220.
\noindent
\bibitem{castellano}     
M.Castellano, N.Cavallo, F.Cevenini, et al., Il Nuovo Cimento, v. 81 B,
No 1, p. 67, 1984.
\noindent
\bibitem{iogansen}     
L.V.Iogansen, M.S.Rabinovich, Sov. Phys. JETP, v.37 (10), 1960, p.83.
\noindent
\bibitem{murphy}     
J.B.Murphy, S.Krinsky, R.L.Glukstern, Proc. of IEEE PAC 1995, Dallas,
(1995).
\noindent
\bibitem{melechin}     
V.N.Melechin, Zhurnal Experim. and Teor. Phys., v.97, No 3 (1990),
p.757.
\noindent
\bibitem{carlsten}     
B.E.Carlsten, Phys. Rev. E, v.54, No 1, (1996), p.838.
\noindent
\bibitem{saldin}     
E.L.Saldin, E.A.Schneidmiller, M.V.Yurkov, Nucl. Instr. Meth., A417, No
4 (1998), p.158.
\noindent
\bibitem{schweizer}     
E.Schweizer, J.Nagel, W.Braun et al., Nucl. Instr. Meth., v.A239
(1985), p. 630.
\noindent
\bibitem{murphy2}     
J.B.Murphy, S.Krinsky, Nucl. Instr. Meth., v.A346 (1994), p.571.
\noindent
\bibitem{hung}     
Hung-chi Lihn, P.Kung, C.Settakorn, H.Wideman, Phys. Rev. Lett., v.76,
No 22 (1996), p.4163.
\noindent
\bibitem{sendai}     
Yu.Shibata, K.Ishi, S.Ono, et al., Phys. Rev. Lett., v.78, No 14
(1997), p.2740; Nucl.  Instr. Meth., v.B145 (1998), p.49.
\noindent
\bibitem{artem}     
Z.L.Artem'eva, E.G.Bessonov, K.N.Shorin, A.S.Yarov, Soviet Physics -
Lebedev Institutes reports, No 1, 1981, p. 28.
\noindent
\bibitem{wartsky}       
L.Wartsky, S.Roland, J.Lasalle, et al., J.
Appl. Phys., v.46, No 8 (1975), p.3644.
\noindent
\bibitem{alex}        
V.I.Alexeev, E.G.Bessonov, A.V.Kalinin, V.A.Krasikov, Preprint FIAN
No 228, Moscow, 1981.
\noindent
\bibitem{al}       
V.I.Alexeev, Ph. D. thesis, Moscow, Lebedev Phys. Inst. RAS, 1989.
\noindent
\bibitem{alferov}        
D.F.Alferov, Yu.A.Bashmakov, K.A.Belovintsev, E.G.Bessonov, P.A.Cherenkov, Phys. - JETP Lett., 1977,
v.26, No 7, p. 385.
\noindent
\bibitem{nikitin}        
M.M.Nikitin, A.F.Medvedev, M.V.Moiseev, Sov. Phys. JETP, v.52, 388, (1980).
\end{thebibliography}
\end{document}